\documentclass[11pt]{article}
\usepackage{times,fancyhdr}
\usepackage{latexsym}
\usepackage{amssymb}
\usepackage{amsmath}
\usepackage{graphicx}
\usepackage{bigfoot}
\usepackage{bm}
\usepackage{sectsty}
\usepackage{dsfont}
\def \d {\mathrm{d}}

\sectionfont{\normalsize}
\subsectionfont{\small}

\newcommand{\htop}{\mathcal{H}_{\mbox{\tiny{top}}}}

\newcommand{\WdW}{Wheeler-DeWitt }
\newcommand{\wdw}{Wheeler-DeWitt }
\newcommand{\kuchar}{Kucha\v{r}}

\newcommand{\fbar}{\overline{f}}
\newcommand{\wtwo}{W_{(2)}}
\newcommand{\zero}{z_{\nu,n}}
\newcommand{\zhalf}{z_{\frac{1}{2},n}}

\date{}

\title{{\normalsize{\bf{TOPOLOGY AND VOLUME EFFECTS IN QUANTUM GRAVITY}}}\\{\small{\bf{WHEELER-D{\footnotesize {E}}WITT THEORY}}}} % Declares the document's title.
\author{{\small Johan Brannlund \footnote{johan.brannlund@concordia.ca}} \\
\it{\small Department of Mathematics and Statistics, Concordia University,} \\
\it{\small Montr\'{e}al, Qu\'{e}bec, Canada, H3G 1M8} \\[-0.1cm]
\line(1,0){45}\\
{\small Andrew DeBenedictis \footnote{adebened@sfu.ca}} \\
\it{\small Department of Physics} \\
\it{{\footnotesize{and}}}\\
\it{\small The Pacific Institute for the Mathematical Sciences,}\\
\it{\small Simon Fraser University, Burnaby, British Columbia, Canada, V5A 1S6}\\[-0.1cm]
\line(1,0){45}\\
{\small Alison Lauman \footnote{aal13@sfu.ca}}\\
\it{\small Department of Physics,} \\
\it{\small Simon Fraser University, Burnaby, British Columbia, Canada, V5A 1S6}}

\begin{document} 

% End of preamble and beginning of text.
\pagestyle{fancy}
\fancyhead{} % clear all header fields
%\fancyhead[EC]{J. Brannlund,\; A. DeBenedictis,\; A. Three}
%\fancyhead[EC]{}
%\fancyhead[EL,OR]{\thepage}
\fancyhead[OR]{\thepage}
\fancyhead[OC]{{\small{TOPOLOGY AND VOLUME EFFECTS IN QUANTUM GRAVITY}}}
\fancyfoot{} % clear all footer fields
\renewcommand\headrulewidth{0.5pt}
\addtolength{\headheight}{2pt} % make space for the rule

\maketitle % Produces the title.
\vspace{-1.0cm}
\begin{center}
\rule{175pt}{0.5pt}
\end{center}

%%begin{abstract}
\begin{center}
 {\footnotesize{\bf{ABSTRACT}}}
\end{center}
\noindent We consider the quantization of space-times which can possess different topologies within a symmetry reduced version of \wdw theory. The quantum states are defined from a natural decomposition as an outer-product of a topological state, dictating the topology of the two-surfaces of the space-time, and a geometric state, which controls the geometry and is comprised of solutions to the \wdw constraints. Within this symmetry reduced theory an eigenvalue equation is derived for the two-volume of spacetime, which for spherical topology is fixed to a value of $4\pi$. However, for the other topologies it is found that the spectrum can be {discrete} and hence the universe, if in one of these other topological states, may only possess certain possible values for the two-volume, whereas classically all values are allowed. We analyze this result in the context of pure gravity (black holes).
\par
\vspace{0.2cm}\noindent\textbf{PACS numbers:} 04.20.Gz\;\;04.60.-m \\
\noindent\textbf{Key words:} topology, quantum gravity, quantum geometry
%%\end{abstract}
\vskip -0.1cm
\vspace{\baselineskip}\hrule

%%%%%%%%%%%%%%%%%%%%%%%%%%%%%%%%%%%%%%%%%%%%%%%
\section{{Introduction}}\label{S:intro}
The quantization of the gravitational field remains one of the most elusive puzzles in modern physics, not least because of the difficulties present when attempting to quantize a theory which is background independent. The realization that general relativity would not lend itself to be satisfactorily quantized via standard techniques dates back at least to the work of Bronstein \cite{ref:bron}. Either quantum mechanics, or general relativity, or both would have to be modified in some way in order to come up with a satisfactory theory of quantum gravitation. An early version of a possible theory of quantized gravity is in the form of \wdw theory \cite{ref:wdw}, which is based on the Arnowitt-Deser-Misner formulation of general relativity \cite{ref:adm}. However, it is now generally accepted that the \wdw theory cannot be a fully correct theory of quantum gravity and that the theory possesses several unresolved problems \cite{ref:landagainst}-\cite{ref:wdwprobssolns}. More recently other more promising background independent theories have been proposed such as Loop Quantum Gravity \cite{ref:LQG1}, Causal Set Theory \cite{ref:CST1}, Causal Dynamical Triangulations \cite{ref:CDT1} etc. These other various approaches eliminate some of the problems present in the \wdw theory.  

Although there are issues with \wdw theory, it is very often used as an approximation to a theory of quantum gravity, and it is therefore useful to see what predictions it makes in this respect \cite{ref:qgeoB}-\cite{ref:qgeo}. Such studies have been performed, for example, in the context of black holes \cite{ref:polbh}-\cite{ref:wdwbh2}, wormholes \cite{ref:visqWH}-\cite{ref:nlemWH}, and cosmology \cite{ref:cookethesis}-\cite{ref:cossing} (also see references therein). With this in mind we perform the analysis here. Given the technical difficulty in dealing with the full theory, we freeze the symmetry first, reducing the number of degrees of freedom at the classical level. The symmetry reduced constraints are then derived via the method of \kuchar\;\cite{ref:kuchar}, and then the constraints are quantized utilizing similar techniques to those used in \cite{ref:kuchar} and \cite{ref:kenmoku}, \cite{ref:kenmoku2}.

Most studies of symmetry reduced quantum gravity concentrate on systems with spherical symmetry, where the two-volume of the constant radius spaces is compact and possesses a value of $4\pi$. However, here we relax this restriction and also allow for two-surfaces with topologies of genus $> 0$. In the higher genus cases the two-volume is not restricted, and appears as a general parameter in the constraint equations which can be set to specific values ``by hand''. In section \ref{S:hilbert} we construct a Hilbert space (formally) which allows the study of various topologies simultaneously. In section \ref{S:geometry} we discuss the geometric sector of the theory, namely the sector governed by \wdw theory. We construct the symmetry reduced constraint equation and look for analogous solutions to those found in \cite{ref:kenmoku} and \cite{ref:kenmoku2}, but for various topologies, which in the geometric sector translates to various two-volumes. In doing so an eigenvalue equation is derived for the two-volume, which leads to a natural definition for an inner product of the geometric eigenstates. It turns out that with appropriate boundary conditions the eigenvalues are \emph{discrete} and therefore the allowable two-volumes is not a continuous arbitrary quantity as is the case in the classical theory. This quantization of geometric quantities is reminiscent of what is found in certain more advanced theories of quantum gravity \cite{ref:avlqg}, \cite{ref:causets}. By studying the specific case of black holes we comment on the classical-quantum correspondence of this result in section \ref{S:solutions}. Finally, we summarize and make some concluding remarks in section \ref{S:conclusion}.

% \subsection{{A subsection}}\label{sch}
% \hskip0.1cm

\section{{The Hilbert Space}}\label{S:hilbert}
The term ``Hilbert space'' is used in the geometric sector in a rather loose sense here, as much of the structure required for a true Hilbert space is not well defined in \wdw theory. One major difficulty is in constructing a physically relevant positive-definite inner product. At this stage, for the geometric sector, we are really simply referring to the space of solutions of the functional differential equation which is the \wdw equation.

We will take the following line element in ADM form for the symmetry reduced systems under consideration:
 \begin{align}
  \d s^{2}=&
\Lambda^2(r,t) (\d r + N^r \d t)^2 + R^2(r,t) \left[ \d \rho^2 + \frac{1}{\beta}
\sinh^2 ( \sqrt{\beta} \rho)  \d \varphi^2 \right]
- N^2(r,t) \d t^2\,,
 \label{eq:line}
 \end{align}
with $0 < \varphi \leq 2\pi$.  $\Lambda(r,t)$ and $R(r,t)$ are the post symmetry reduction configuration degrees of freedom and correspond to the metric component $g_{rr}$, and the analog of the Tolman-Bondi 2-surface conformal factor respectively. The constant $\beta$ controls the possible topologies of the space-time's 2D subspaces. The allowed topologies are as follows:\\
i) $\beta=-1$: In this case $(\rho,\,\varphi)$ sub-manifolds are spheres. \\
ii) $\beta=0$: In this case $(\rho,\,\varphi)$ sub-manifolds are tori and these surfaces for this case are intrinsically flat. \\
iii) $\beta=1$: In this case $(\rho,\,\varphi)$ sub-manifolds are surfaces of constant negative curvature of genus $g > 1$, depending on the identifications chosen. Such surfaces may be compact or non-compact \cite{ref:complexteich}, \cite{ref:geo3} (also, please refer to the appendix for details).

\noindent Furthermore, for simplicity, we consider only the pure gravitational sector of the theory, as adding even simple matter to the system results in an extremely complicated scenario when considering quantization. \\

We wish to treat the whole set of topologies simultaneously, instead of each one individually, as we do not consider the universe to be in any particular topology eigenstate. We therefore require a Hilbert space which allows for this and a consistent way of achieving this is via a tensor product space of the form
\begin{equation}
 \mathcal{H}=\mathcal{H}_{\mbox{\tiny{WdW}}} \otimes \mathcal{H}_{\mbox{\tiny{top}}}\,, \label{eq:hilbert}
\end{equation}
where $\mathcal{H}_{\mbox{\tiny{WdW}}}$ is some Hilbert space of usual Wheeler-DeWitt theory (which is technically not well defined at this stage), with some form of Wheeler-DeWitt inner product (which is also not technically not well defined at this stage), and $\mathcal{H}_{\mbox{\tiny{top}}}$ is the ``topological sector'' of the Hilbert space. Hence we have rays which are both geometric (containing information related to geometry) and topological (containing information related to topology). The \WdW Hilbert space is the space of all 3-metrics, $q_{ab}$, which are subject to the symmetry reduction above. That is, it is the space of all 3-metrics with 2-D subspaces given by the sub-element of (\ref{eq:line}). The purpose of the topological Hilbert space is to constrain these 2-D subspace to specific topological values. That is, when the wave-function is in an eigenstate corresponding to some specific topology, then the metrics must only be the ones compatible with this topology. Another way to put this is that when the universe is in a topological eigenstate the metrics considered must be symmetry reduced to be compatible with that topology. Therefore, the eigenstates of topology, which span $\htop$, are trivially of the form
\begin{equation}
 \left|\beta_{n}\right>=\sum^{1}_{\beta=-1}\left|\beta\right>\left<\beta\right|\left.\beta_{n}\right>=\sum^{1}_{\beta=-1}\left|\beta\right>\delta_{\beta\beta_{n}}\,, \label{eq:topeig}
\end{equation}
where $\beta_{n}$ is one of the three possible values $\beta$ can possess (-1, 0, +1). This is a discrete quantum number and hence, as expected, is a topological quantum number. When the system is in a topology eigenstate (the $n^{\mbox{\tiny{th}}}$ eigenstate), we have the following:
\begin{equation}
 \left|\Psi_{n}\right>=\left|\psi\left(q_{ab}\left(\beta\right)\right)\right> \otimes \left|\beta_{n}\right> =
\!\!\left\{\!\!\!
\begin{array}{l}
\;\;\left|\psi\left(q_{ab}\left(\beta_{n}\right)\right)\right> \otimes \left|\beta_{n}\right>\;\;\mbox{if}\;\:\beta=\beta_{n}, \\[0.2cm]
\qquad\qquad 0\qquad\qquad\quad\;\mbox{if}\;\, \beta\neq \beta_{n}
\end{array} \right.\;\;, \label{eq:eigenstate}
\end{equation}
which is compatible with the requirements stated above.

The topological eigenstates satisfy the condition
\begin{equation}
 \left<\beta_{n}|\beta_{m}\right> =\sum_{\beta}\delta_{\beta_{n}\beta}\,\delta_{\beta\beta_{m}} =\delta_{\beta_{n}\beta_{m}}\,
\end{equation}
and therefore states of different topology are orthogonal. Among other things, this implies that if the universe is in a state of a certain topology, it cannot spontaneously change its topology. We have no mechanism at this stage to implement the possibility of a dynamic topology. Strictly speaking, topology is a measurable quantity, in that an observer can theoretically travel in some direction in the universe and ``measure'' whether or not they eventually come back to their spatial starting point along certain trajectories, hence measuring the topology. Therefore it is not surprising that the topology spectrum is real and that the topology eigenfunctions are orthogonal. It may be interesting to attempt to extend the Hilbert space so that the genus number for the $\beta=1$ case plays a role, perhaps via the utilization of non-homotopic chains wrapping the handles as degrees of freedom. This avenue is not pursued here. 

\section{The Geometric Sector}\label{S:geometry}
Here we concentrate on the geometry; namely the \WdW sector of the theory, which involves finding solutions to the symmetry reduced quantum Hamiltonian and diffeomorphism constraints. We will generalize the approach of \cite{ref:kenmoku} to the metrics in
equation (\ref{eq:line}).

A lengthy calculation yields the classical Hamiltonian and diffeomorphism constraints governing the geometry, which are given by\footnote{The calculations to derive equations (\ref{eq:symham}) and (\ref{eq:symdif}) follow somewhat the method in \cite{ref:kuchar} and \cite{ref:kenmoku} for spherical symmetry. However, the calculations need to be done from scratch as the 2-volume, $\wtwo $, and the topological parameter, $\beta$, are present here, and it is not a priori possible to discern how they enter into the constraints.}
\begin{subequations}
\begin{align}
  H &= \frac{1}{2} \frac{p_\Lambda^2 \Lambda}{{\wtwo} R^2} - 
\frac{p_\Lambda p_R}{{\wtwo} R} + {\wtwo} \frac{RR''}{\Lambda} - 
{\wtwo} \frac{R \Lambda' R'}{\Lambda^2} + \frac{{\wtwo}R'^2}{2\Lambda} +
\frac{{\wtwo} \Lambda \beta}{2} + \frac{{\wtwo} \lambda \Lambda R^2}{2}\,, \label{eq:symham}
\\[0.2cm]
H_r &= p_R R' - p_\Lambda' \Lambda\,. \label{eq:symdif}
\end{align}
\end{subequations}
Here $\wtwo =V_2/4\pi$ is the normalized two-volume of the submanifolds coordinatized
by\footnote{In general the area integral for the compact 2-surface has an upper-limit along some curve given by $\rho=\varrho\hspace{-0.45mm}(\varphi)$. In the spherical case $\rho$ is simply the polar angle and hence $\varrho\hspace{-0.45mm}(\varphi)=\mbox{constant}=\pi$, and $V_{(2)}=4\pi$. In the non spherical cases, the coordinate $\rho$ is a radial coordinate on the 2-surface and $V_{(2)}$ can take on arbitrary values. (See the appendix for details.)} $\rho$ and $\varphi$, and $p_R$ and $p_\Lambda$ are the momenta conjugate
to the configuration space variables $R$ and $\Lambda$. The quantity $\lambda$ represents the cosmological constant, which classically can have any value for the spherical case, but must be negative for the other scenarios \cite{ref:aminneborg}-\cite{ref:lemos}.

Passing to the quantum regime, we promote the above constraints to operators,
resulting in
\begin{subequations}
\begin{align}
 \hat{H} &= \frac{1}{2{\wtwo}} \Lambda R^{-2} \hat{p}_\Lambda^{(C)} 
\hat{p}_\Lambda -
\frac{R^{-1}}{{\wtwo}} \hat{p}_R \Lambda \hat{p}_\Lambda^{(B)} \Lambda^{-1} +
{\wtwo} \Lambda R'^{-1} \left( \frac{R}{2} (\chi - F )\right)'\,, \label{eq:Qsymham} \\[0.2cm]
\hat{H}_R &= R' \hat{p}_R - \Lambda \hat{p}'_\Lambda\,, \label{eq:Qsymdif}
\end{align}
\end{subequations}
where primes denote partial derivatives with respect to $r$ and
\begin{align}
  \chi &:= \Lambda^{-2} R'^{2}\,, \nonumber \\
   F &:= - \beta - 2mR^{-1} - \frac{\lambda}{3}R^2\,. \label{eq:chiandF}
\end{align}
This form of $F$, if $m$ is constant, comes from considering vacuum solutions (which we do here). Later will will concentrate specifically on the case of black hole space-times.
In (\ref{eq:Qsymham}) and (\ref{eq:Qsymdif}) $\hat{p}_\Lambda$ and $\hat{p}_R$ are the (functional) Schr\"{o}dinger momentum operators; $\hat{p}_\Lambda=-i\frac{\delta}{\delta \Lambda(r)}$ and $\hat{p}_R=-i\frac{\delta}{\delta R(r)}$ respectively. (The $t$ dependence is henceforth dropped due to the equal time nature of the quantization.) 

The quantities $\hat{p}_\Lambda^{(A)}$, $\hat{p}_\Lambda^{(C)}$ and $\hat{p}_\Lambda^{(B)}$ 
are given by 

\begin{align}
  \hat{p}_\Lambda^{(A)} &= A \hat{p}_\Lambda A^{-1}
 \nonumber  \\
\hat{p}_\Lambda^{(B)} &=  A^{1/2} \hat{p}_\Lambda A^{-1/2}
=\frac 1 2 \left(   \hat{p}_\Lambda +   \hat{p}_\Lambda^{(A)}\right)
 \nonumber \\
\hat{p}_\Lambda^{(C)}&=  C \hat{p}_\Lambda C^{-1} =
\hat{p}_\Lambda^{(A)} - i R R'^{-1} \left( A^{-1} 
\frac{\delta A } {\delta \Lambda} \right)'
\end{align}

where

\begin{displaymath}
  C=A\exp\left( - \int R R'^{-1} \int \left( A^{-1} 
\frac{\delta A } {\delta \Lambda} \right)' \d \Lambda \, \d r \right)
\end{displaymath}
Note that $A$ and $C$ are ordering functions which yield a similarity transformation on the operators they act with. The form of the quantum constraints is dictated by imposing the reduction
to the classical constraints when operator ordering is ignored, as well as
ensuring that equation (\ref{massop}) holds classically as well as
quantum-mechanically. 

To acquire analytic solutions it was noted in \cite{ref:kenmoku} that it is useful to define the quantity $Z$ by
\begin{equation}
  Z = \int \d r \Lambda f(R,\chi) = \int dr \int d\Lambda\, \overline{f}(R,\,\chi)\,\,, \label{eq:Z}
\end{equation}
where $f$ and $\overline{f}$ are arbitrary functions. The importance of $Z$ lies in the fact
that it commutes with the diffeomorphism constraint $\hat{H}_r$: 
$[Z,\hat{H}_r]=0$. Therefore, solutions $\Psi$ of the diffeomorphism constraint
$\hat{H}_r \Psi=0$ will only depend on $Z$. 

Assuming $\Psi$ to be only $Z$-dependent as above, it is then required to be a 
solution of the quantum Hamiltonian constraint $\hat{H}\Psi=0$ (the
\WdW equation). This will be done in a somewhat roundabout way
by introducing the quantum mass operator $\hat{M}$, defined by 

\begin{displaymath}
  \hat{M}-m = \frac{1}{2\wtwo} R^{-1} \hat{p}_\Lambda^{(A)} \hat{p}_\Lambda
- \frac{\wtwo}{2} R ( \chi - F)\,
\end{displaymath}
($m$ is an eigenvalue of $\hat{M}$).

Similarly to the remarks made above, the form of this operator is
dictated by reduction to the classical form when operator ordering is ignored.
%% See \cite{ref:kenmoku} for details.

Most important for the study here is that $\hat{M}$ obeys the relation
\begin{equation}
\label{massop}
  \hat{M}' = - \Lambda^{-1} R' \hat{H} - R^{-1} \hat{p}_\Lambda^{(B)}
\Lambda^{-1} \hat{H}_r\,.
\end{equation}
This means that if $\Psi$ is a solution of the diffeomorphism constraint
$\hat{H}_r\Psi=0$ and the mass constraint $\hat{M}\Psi=m\Psi$, it is
also a solution of the \wdw equation $\hat{H}\Psi=0$. 
Proof: By taking a derivative with respect to $r$, we get that
$\hat{M}\Psi=m\Psi \Rightarrow \hat{M}' \Psi = 0$. The crucial conceptual
aspect of this derivation is that $\Psi$ is a functional and therefore
does not depend on $r$.

Now, writing out the condition $(\hat{M}-m)\Psi=0$ leads to 
\begin{equation}
\left(  \frac{\delta Z}{\delta \Lambda} \right)^2 \frac{\d^2 \Psi}{\d Z^2}
+A \left[ \frac{\delta }{\delta \Lambda} \left( A^{-1}
\frac{\delta Z}{\delta \Lambda} \right) \right] \frac{\d \Psi}{\d Z}
+ \wtwo^{2} R^2(\chi-F) \Psi = 0\,. \label{eq:aftermasswfw}
\end{equation}

At this point it is useful to choose $A=A_Z(Z)\bar{A}(R,\chi)$ and
\begin{displaymath}
  \bar{A}(R,\chi)=\frac{\delta Z}{\delta \Lambda} = R\sqrt{\chi-F}\,,
\end{displaymath}
which leads to 
\begin{equation}
 \frac{\d^2 \Psi}{\d Z^2}- A_Z^{-1} \frac{d A_Z}{d Z}
 \frac{\d \Psi}{\d Z} + \wtwo^{2} \Psi = 0\,. \label{eq:simpleWdW}
\end{equation}
Although the above equation does not explicitly depend on the topological parameter $\beta$, we will show below how $\beta$ enters in specific solutions. Also of interest is the appearance of the normalized two-volume in the last term, indicating that non-trivial volume effects will be present in the solutions.

\section{Solutions}\label{S:solutions}
In the spherical case, an analytic solution in the form of Bessel functions was discovered \cite{ref:kenmoku} \cite{ref:kenmoku2}. Here we attempt to find an analogous solution. Consider the choice
\begin{equation}
 A_{Z}=Z^{2\nu-1}\,, \nonumber
\end{equation}
which transforms (\ref{eq:simpleWdW}) to
\begin{equation}
  \frac{\d^2 \Psi}{\d Z^2}-(2\nu-1)Z^{-1}\frac{\d \Psi}{\d Z} + \wtwo^{2} \Psi = 0\,. \label{eq:besseleq}
\end{equation}
The solutions to this second order equation are Bessel functions of the first and second kind:
\begin{equation}
 \Psi(Z)=C_{1}\, Z^{\nu}\, J_{\nu}(\wtwo Z) + C_{2}\, Z^{\nu}\, Y_{\nu}(\wtwo Z)\,, \label{eq:besselsol}
\end{equation}
where $J_{\nu}$ and $Y_{\nu}$ are Bessel functions of the first and second kind respectively and $C_{1}$ and $C_{2}$ are constants. It is interesting to note that the normalized two-volume, $\wtwo $ appears as a frequency and therefore not only controls the number of oscillations, but also to some extent the rate at which the solutions fall off or grow as a function of $Z$. Recall that in the case of spherical topology $\wtwo $ is fixed to unity, whereas in the other scenarios $\wtwo $ could be arbitrarily large.

It can be noted that the equation (\ref{eq:besseleq}), with boundary conditions, is an eigenvalue equation for $\wtwo $, and hence the solutions (\ref{eq:besselsol}) represent the eigenfunctions for the normalized two-volume. Although there is no universally accepted inner-product for \wdw theory, we can exploit here the fact that in principle the 2-volume is a measurable observable, and its operator, as given by the first two terms in (\ref{eq:simpleWdW}) or (\ref{eq:besseleq}), must be Hermitian since $\wtwo$ is a real quantity\footnote{The non-locality of this quantity is expected since in the symmetry reduced theory the 2-volume is a conserved quantity and any quantity which Poisson commutes with the constraints of general relativity is non local \cite{ref:torre}. See also \cite{ref:kieferbook}.}. Hence the eigenfunctions corresponding to different eigenvalues must be orthogonal. This orthogonality requirement allows us to fix an acceptable inner-product as there exists only the following natural orthogonality relationship for these Bessel functions:
\begin{subequations}
\begin{align} 
\int_{0}^{b} J_{\nu}\left(\frac{{\zero}\,Z}{b}\right)\,J_{\nu}\left(\frac{z_{\nu,n^{\prime}}Z}{b}\right)Z\,dZ =& \delta_{n,n^{\prime}} \frac{b^{2}}{2} J^{2}_{\nu+1}({\zero})\,, \label{eq:orthog1}\\
\int_{0}^{\infty} J_{\nu}(kZ)\,J_{\nu}(k^{\prime}Z)Z\,dZ = & \frac{1}{k}\delta(k-k^{\prime})\,. \label{eq:orthog2}
\end{align}
\end{subequations}
Here,  ${\zero}$ represents the $n^{\mbox{\footnotesize{th}}}$ zero of the Bessel function $J_{\nu}(\cdot\cdot)$. The expressions (\ref{eq:orthog1}) and (\ref{eq:orthog2}) correspond to finite and infinite intervals respectively. Note that this requires us to take $C_{2}=0$ in (\ref{eq:besselsol}) and limits the acceptable solutions (\ref{eq:besselsol}) to Bessel functions of order $\nu=1/2$ (trigonometric). This result also eliminates unbounded wave functions since for large $Z$, $J_{\nu}(\wtwo Z) \sim Z^{-1/2}$ so our solutions, which are of the form $Z^{\nu} J_{\nu}(\cdot\cdot)$, remain bounded at large $Z$. (This is especially useful for the infinite interval case (\ref{eq:orthog2}).) Furthermore, an interesting result that arises is that if the upper-limit of integration in (\ref{eq:orthog1}) is fixed (for example, if it is dictated by a Dirichlet boundary condition), then the spectrum of two-volumes is \emph{discrete}. In the finite interval case we also require that the wave functions be normalized to unity which, using (\ref{eq:orthog1}), yields the condition
\begin{equation}
|C_{1}|^{2} \int_{0}^{b} J_{\frac{1}{2}}^{2}\left(\wtwo Z\right)Z\,dZ = |C_{1}|^{2} \frac{b^{2}}{2} J^{2}_{\frac{3}{2}}(\wtwo  b) = 1\,, \label{eq:normalization}
\end{equation}
and therefore the constant $C_{1}$ is set via the relation
\begin{equation}
|C_{1}|^{2} = \frac{2}{b^{2}}J^{-2}_{\frac{3}{2}}(\wtwo b)\,. \label{eq:normalizationconst}
\end{equation}
(Note that $\wtwo ={{\zhalf}}/{b}$ where ${\zhalf}$ is a zero of $J_{\frac{1}{2}}$, \emph{not of} $J_{\frac{3}{2}}$, and hence $|C_{1}|^{2}$ is well defined.) For the case of spherical topology the normalized two-volume must have the value $\wtwo =1$. In the other cases, it is possible to have different but discretely quantized values of $\wtwo$.

The parameter controlling the allowable topologies, $\beta$, is not explicitly present in the above solution. We therefore wish to reintroduce this parameter. First, we consider changing the coordinates to a form more suitable for studying the ``$T$-domain'' of a black hole (the time-dependent interior). Therefore, we rewrite line element (\ref{eq:line}) as follows (setting the shift vector to zero now since we have already derived the equations of motion):
\begin{align}
  \d s^{2}=&
\Lambda^2(x,\tau)\, \d x^2 + T^2(x,\tau) \left[ \d \rho^2 + \frac{1}{\beta}
\sinh^2 ( \sqrt{\beta} \rho)  \d \varphi^2 \right]
- N^2(x,\tau)\, \d \tau^2\,.
 \label{eq:tline}
 \end{align}
Switching the notation of the configuration variable from $R$ to $T$ is to reflect the fact that in the interior of a black hole the ``radial'' coordinate becomes time-like. $x$ is the $T$-domain spatial coordinate which corresponds to the coordinate $t$ in the exterior of the black hole (the ``$R$-domain'') and $\tau$ is the interior time which corresponds to the exterior radial coordinate $r$.

Consider the choice
\begin{equation}
 \fbar=T\sqrt{\chi-F(T)}\,,
\end{equation}
where $\chi$ and $F(T)$ were defined in (\ref{eq:chiandF}) with the change $R\rightarrow T$ here. This yields, via (\ref{eq:Z}):
\begin{equation}
 Z=\int dx \int d\Lambda\, T \sqrt{\chi-F}\,.
\end{equation}
Next set $T^{\prime}=0$, and $\dot{T}=1$ where the prime denotes differentiation with respect to $x$ and the dot with respect to $\tau$. Then $\chi=0$ and
\begin{equation}
 Z=\int dx\,\Lambda T \sqrt{-F}\,. \label{eq:zxint}
\end{equation}
However, in this case $\Lambda=\sqrt{-F}$ so
\begin{equation}
 Z=-\int_{0}^{x_{0}}dx\,TF = -TFx_{0}\,,
\end{equation}
and hence
\begin{equation}
 Z=\left(\beta T +2m +\frac{\lambda}{3} T^{3}\right) x_{0}\,. \label{eq:ZofT}
\end{equation}
The wave function $\Psi(Z)$ has now been converted to $\Psi(T)$ and the $\beta$ parameter is now explicitly present. 

Note that at this stage the symmetry is now completely frozen, and one has effectively turned the theory into one reminiscent of standard quantum mechanics. The new residual degree of freedom is now the variable $T$. In this case these black hole wave functions take the form
\begin{equation}
\Psi(T)=C_{1}\, \left(\beta T +2m +\frac{\lambda}{3} T^{3}\right)^{\frac{1}{2}} x_{0}^{\frac{1}{2}}\: J_{\frac{1}{2}}\left(\wtwo \left(\beta T +2m +\frac{\lambda}{3} T^{3}\right) x_{0}\right)\,. \label{eq:besselT}
\end{equation}
The constant $b$ is set by the condition that now $Z=b \Rightarrow \left(\beta T_{b} +2m +{\lambda}/{3}\, T_{b}^{3}\right) x_{0} = b$, with $T_{b}$ being the specific value of $T$ when this relation holds. To set the parameter $b$ note that classically the domain of validity for the solutions is in the range $0 < T < T_{h}$, with $T_{h}$ being the horizon value of $T$ for the black hole (that is, the positive real root of  (\ref{eq:ZofT})). $Z=0$ corresponds to the upper limit, $T=T_{h}$. Therefore $Z=b$ corresponds to the other end of the domain of validity, namely $T=0$. Therefore, setting $T=0$ in (\ref{eq:ZofT}) and noting that this must equal $b$ yields the condition
\begin{equation}
 b=2mx_{0}. \label{eq:bvalue}
\end{equation}
We now have a well-posed Dirichlet problem with $T=0$ and $T=T_{h}$ as the boundary surfaces. 

The interpretation of the wave function here is as follows. The probability density $\Psi^{\dagger}(T)\Psi(T)$ yields a measure of the probability that, for that value of $T$, the metric component $\Lambda^{2}(T)$ corresponds to its classical value of $\Lambda^{2}_{\mbox{\tiny{class}}}(T)=-F(T)= -(- \beta - 2mT^{-1} - \frac{\lambda}{3}T^2)$. That is, it is the probability that the metric's form is the one that it has been reduced to.

\begin{figure}[h!t!]
\begin{center}
\includegraphics[bb=9 0 743 312, clip, scale=0.552,keepaspectratio=true]{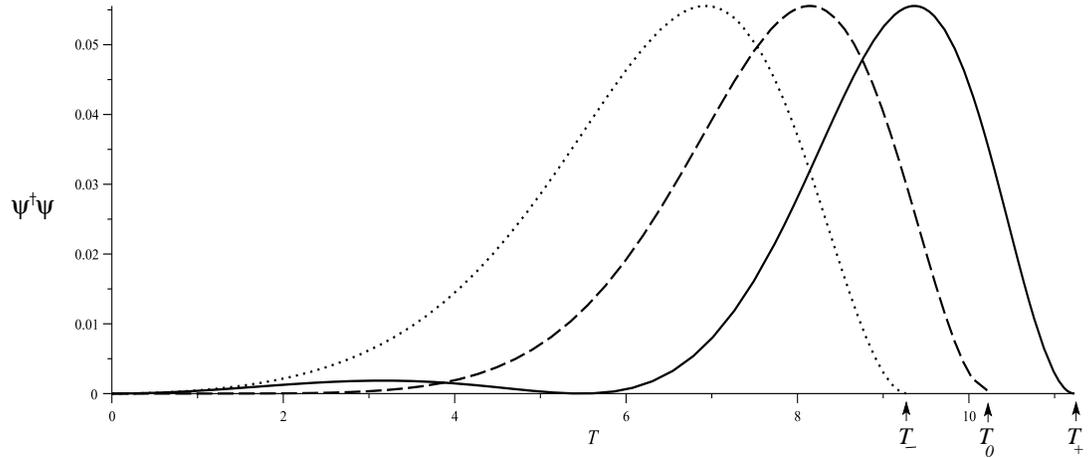}
\caption{\small{The normalized probability density for the $n=1$ mode. The genus $>1$ case ($\beta=+1$) is solid, the genus = 1 case ($\beta=0$) is dashed, and the genus $=0$ case ($\beta=-1$) is dotted. The values are as follows: $m=18$, $x_{0}=1$, $\lambda=-0.1$. The $T$ range for the three cases differ due to the different location of the horizon for the three cases ($T_{+}$, $T_{0}$, $T_{-}$ respectively).\vspace{0.5cm}}}\label{fig:f1}
\end{center}
\end{figure}

\begin{figure}[h!t!]
\begin{center}
\includegraphics[bb=9 0 743 312, clip, scale=0.552,keepaspectratio=true]{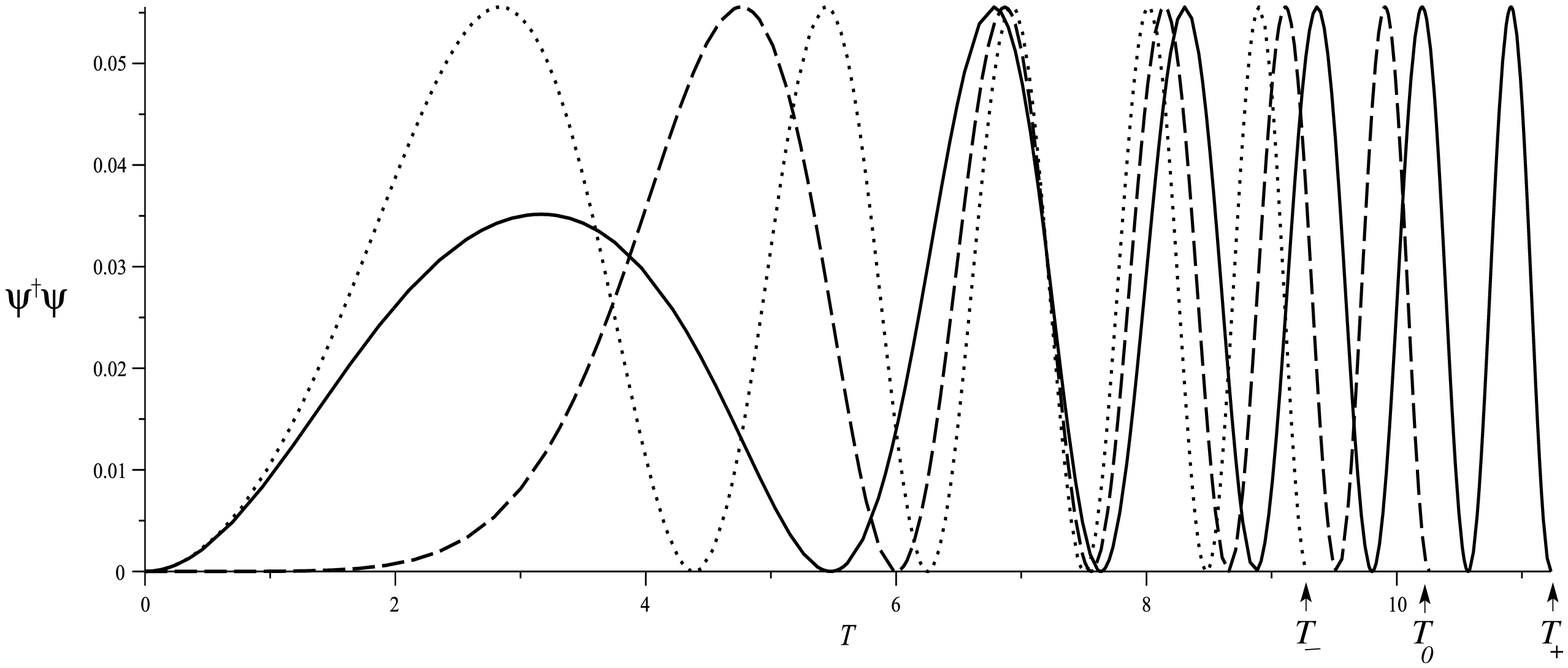}
\caption{\small{The normalized probability density for the $n=5$ mode. The genus $>1$ case ($\beta=+1$) is solid, the genus = 1 case ($\beta=0$) is dashed, and the genus $=0$ case ($\beta=-1$) is dotted. The values are as follows: $m=18$, $x_{0}=1$, $\lambda=-0.1$. The $T$ range for the three cases differ due to the different location of the horizon for the three cases ($T_{+}$, $T_{0}$, $T_{-}$ respectively).\vspace{0.5cm}}}\label{fig:f2}
\end{center}
\end{figure}

\begin{figure}[h!t!]
\begin{center}
\includegraphics[bb=9 0 743 312, clip, scale=0.552,keepaspectratio=true]{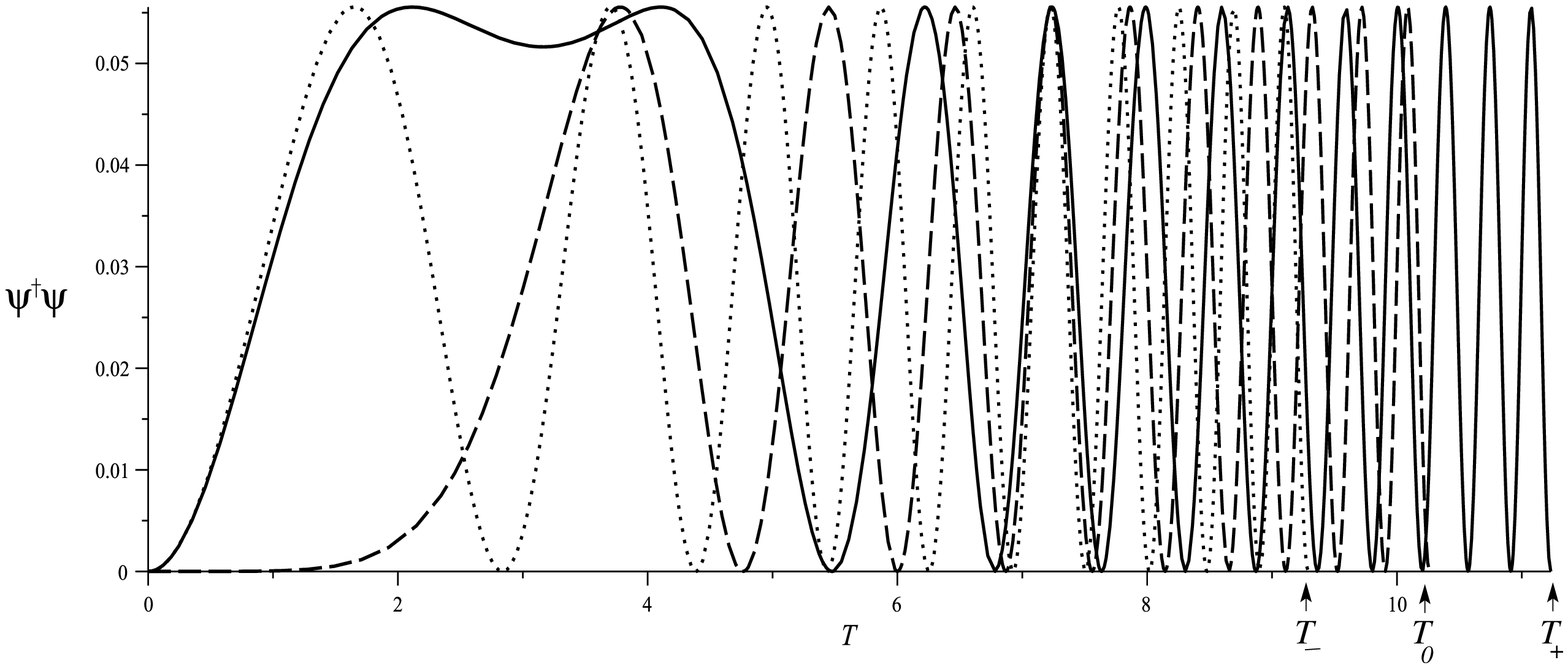}
\caption{\small{The normalized probability density for the $n=10$ mode. The genus $>1$ case ($\beta=+1$) is solid, the genus = 1 case ($\beta=0$) is dashed, and the genus $=0$ case ($\beta=-1$) is dotted. The values are as follows: $m=18$, $x_{0}=1$, $\lambda=-0.1$. The $T$ range for the three cases differ due to the different location of the horizon for the three cases ($T_{+}$, $T_{0}$, $T_{-}$ respectively).\vspace{0.5cm}}}\label{fig:f3}
\end{center}
\end{figure}

As mentioned previously, the zeroes of the Bessel function introduce a quantum number, which is denoted here as $n$ and is defined via:
\begin{equation}
\zhalf=n\pi=\wtwo b = 2 m x_{0} \wtwo\,. \label{eq:zerorelations}
\end{equation}
Different values of $n$ yield different modes and also different possible values for $\wtwo$ in the case of $\beta=+1$ and $\beta=0$. For the spherical case ($\beta=-1$), $\wtwo$ must equal $1$ and therefore (\ref{eq:zerorelations}) seems to imply that the mass must be quantized, or else the spatial ``size'' of the domain considered inside the black hole must be quantized, or some combination of both. However, since the integral over the spatial slice inside the black hole in (\ref{eq:zxint}) is arbitrary, quantized masses are the more likely result. It is interesting that quantization of spherical black hole mass in \wdw theory and related methods has been noted previously in \cite{ref:masskat}, \cite{ref:vazwitten} via a different approach.  We plot several modes for the various topologies simultaneously in figures \ref{fig:f1}-\ref{fig:f3}.

As expected, the higher the value of $n$, the more oscillations are present in the domain. This yields a quantum-classical correspondence at high $n$. That is, since any measuring device will have a finite resolution, at very high $n$ one would measure the classical
value of the metric, $\Lambda^{2}(T)=\Lambda^{2}_{\mbox{\tiny{class}}}(T)$ (or, more strictly speaking, the corresponding orthonormal Riemann
tensor, which is measurable via tidal forces) with equal probability at all values of $T$ and therefore at high $n$ the classical picture emerges. This is analogous to the situation of confined particles in ordinary quantum mechanics, where classical probability measurements emerge at large values of the quantum numbers due to the presence of a higher frequency in the wave function. It is interesting to note from the solutions that near $T=0$ this effect is less pronounced. This is perhaps not surprising as one expects quantum gravitational effects to deviate more strongly from their classical counter-parts in the high curvature region near the singularity ($T=0$). Interestingly, the metrics compatible with toroidal topology are the ones that behave least classically towards $T=0$ even for moderately large $n$.

\section{Concluding Remarks}\label{S:conclusion}
A symmetry reduced version of \wdw geometrodynamics was utilized to study quantum gravity effects in space-times compatible with different topologies. A Hilbert space was constructed consisting of a topological sector and a geometric sector, the latter being the space of solutions to the symmetry reduced \wdw constraints. An eigenvalue equation was derived for the normalized two-volumes of the space-times which allows us to construct a unique inner-product for the eigenstates, and therefore normalize the eigenfunctions. It is found that with Dirichlet boundary conditions the two-volume possesses a discrete spectrum, and thus the observed universe may not possess an arbitrary value of the two-volume, unlike in the classical case. This aspect is controlled by a quantum number $n$ and was analyzed in detail in the context of black holes. From the form of the eigenfunctions for large values of $n$ it is expected that the classical value of the metric is measured with equal probability, and hence one has a sort of quantum-classical correspondence at large quantum number. The classical behavior at large $n$ is less pronounced near the black hole singularity, which is taken as an indicator that quantum gravity effects are more important in high curvature regions. Out of the topologies considered, the metrics compatible with genus 1 (toroidal) possess the least classical behavior.

\section*{Acknowledgments}
JB was partially funded by the Atlantic Association for Research in the Mathematical Sciences (AARMS).
AD is grateful for the kind hospitality of the Department of Mathematics and Statistics, Concordia University, Montr\'{e}al, where part of this work was carried out. This was made possible by an AARMS travel grant.

\appendix
\numberwithin{equation}{section}
\section{Appendix: The construction of the 2-surfaces}

In this appendix we briefly overview the structure of the 2-surfaces for the various values of $\beta$.

\subsection{$\bm{\beta=-1}$}

In this case the 2-surface line element of (\ref{eq:line}) becomes:
\begin{equation}
d\sigma_{(2)}^{2}=\d\rho^{2}+\sin^2 (\rho)  \d \varphi^2\,, \label{eq:sphereline}
\end{equation}
which is the line-element on a 2-sphere and possesses a two-dimensional volume of
\begin{equation}
 V_{(2)}=4\pi\wtwo=\int_{0}^{2\pi}\int_{0}^{\pi} \sin(\rho)\,\d\rho\,\d\varphi = 4\pi\,,
\end{equation}
where $\wtwo$ is the normalized 2-volume which we often refer to as simply the 2-volume. The coordinates must span the above range in order for the chart to have full, but not multiple, coverage (save for the usual polar point) of the 2-sphere and so that the 2-surface is geodesically complete.

\subsection{$\bm{\beta=0}$}

\begin{figure}[h!t!]
\begin{center}
\includegraphics[bb=0 0 275 325, clip, scale=0.5,keepaspectratio=true]{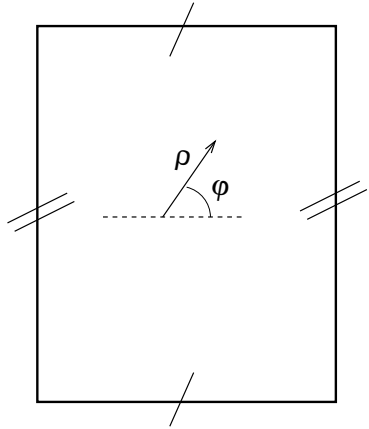}
\caption{\small{The creation of a torus via the identification of a flat space. Note that the 2-volume of the torus can be arbitrary, as the length and width of the surface is not constrained.\vspace{0.5cm}}}\label{fig:torus}
\end{center}
\end{figure}

In this case the 2-surface line element of (\ref{eq:line}) becomes:
\begin{equation}
d\sigma_{(2)}^{2}=\d\rho^{2} + \rho^{2}\d\varphi^{2} \,, \label{eq:torline}
\end{equation}
which is the line-element on a flat 2-surface in polar coordinates. Toroidal black hole space-times are constructed via the identification show in figure \ref{fig:torus}

Note that in this case, the condition of geodesic completeness does not constrain the range of the $\rho$ coordinate (hence the size of the plane is unspecified) and therefore the 2-volume can be of any size.

\subsection{$\bm{\beta=+1}$}

This case is arguably the most difficult one to visualize. As before we begin with the 2-surface line element of (\ref{eq:line}), which for $\beta=+1$ becomes:
\begin{equation}
d\sigma_{(2)}^{2}=\d\rho^{2} + \sinh^2 (\rho) \d \varphi^2\,. \label{eq:genusline}
\end{equation}
To show that this is the line-element of a hyperbolic plane of constant curvature, which can be identified to create surfaces of arbitrary genus, we begin with the following coordinate transformation:
\begin{subequations}
 \begin{align}
\rho=&\sinh^{-1}(\sqrt{|k|}\tilde{r})\,,\label{eq:transf1}\\
d\rho=&\frac{\sqrt{|k|}}{\sqrt{1+|k|\tilde{r}^{2}}}\,\d \tilde{r}\,,\label{eq:transf2}
\end{align}
\end{subequations}
with $k <0$ a constant. Using this, the line-element (\ref{eq:genusline}) may be re-written as:
\begin{equation}
d\sigma_{(2)}^{2}=|k|\left[\frac{\d \tilde{r}^{2}}{\sqrt{1-k\tilde{r}^{2}}}+\tilde{r}^{2}\d\varphi^{2}\right]\, , \label{eq:flrwline}
\end{equation}
which is the well-known Friedmann-Lema\^{i}tre-Robertson-Walker line-element for constant negative curvature (recall $k<0$). Such a plane of negative curvature is compatible with topologies of genus $>1$. We briefly review the identification here along the lines found in, for example, \cite{ref:nakahara}.

\begin{figure}[!h!]
\begin{center}
\includegraphics[bb=0 0 1125 465, clip, scale=0.35,keepaspectratio=true]{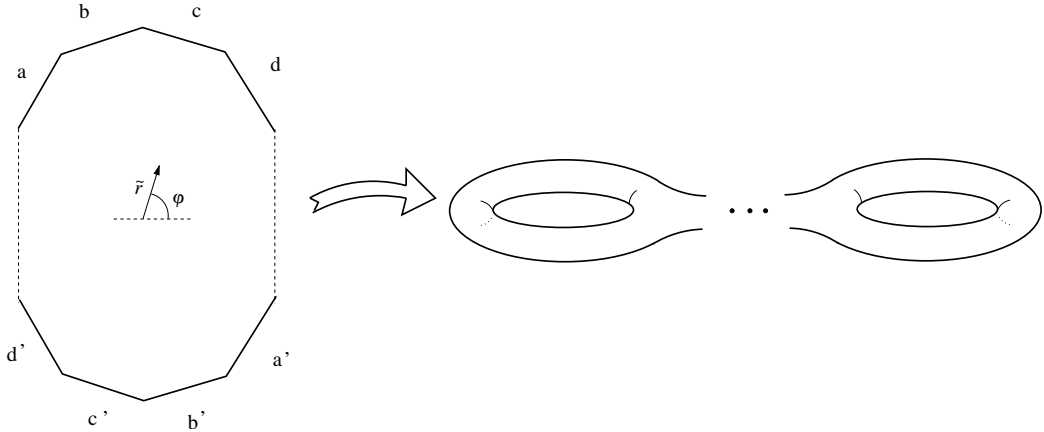}
\caption{\small{The creation of higher genus surface via the identification of a hyperbolic plane. Note that the 2-volume of the surface can be arbitrary, as the span of $\tilde{r}$ to the identification lines is not constrained by any condition. That is, the polygon can be of any size. \vspace{-0.5cm}}}\label{fig:genus}
\end{center}
\end{figure}

In the left of figure \ref{fig:genus} a \emph{qualitative idealization} of the hyperbolic plane is shown. The ``line segments'' (in reality not necessarily straight lines) labeled a, b, c, d, shown  are to represent the ``lines'' along which the identification to the corresponding ``lines'' $\mbox{a}^{\prime},\, \mbox{b}^{\prime},\,\mbox{c}^{\prime},\,\mbox{and }\mbox{d}^{\prime}$ are made. The dashed lines in the left figure represent other possible line segments that can be identified (the constraint being that there must be $4g$ sides to the polygon in total to create a genus $g$ identification). The diagram on the right represents a qualitative representation of the surface created via the identifications. The two handles shown are created via the identifications $\mbox{a}\leftrightarrow\mbox{a}^{\prime},\, \mbox{b}\leftrightarrow\mbox{b}^{\prime},\, \mbox{c}\leftrightarrow\mbox{c}^{\prime},\, \mbox{and }\mbox{d}\leftrightarrow\mbox{d}^{\prime}$. The dots represent other possible handles via the identification of more sides of the polygon (as represented by the dashed lines on the left diagram). Note that again in this case, the condition of geodesic completeness does not constrain the range of the coordinate $\tilde{r}$ and therefore the 2-volume can be of any size.

%% \newpage
\vspace{0.5cm}
\renewcommand{\refname}{{References}}
\linespread{0.6}
\bibliographystyle{unsrt}

\end{document}